\documentclass[12pt]{article}

\begin{document}
\pagestyle{empty}
\begin{center}
{\Large \bf Is it possible 
to check urgently   
the 5-loop  analytical results for the  $e^+e^-$-annihilation 
Adler function ?} 

\vspace{0.1cm}

{A.L. Kataev }\\
\vspace{0.1cm}

Institute for Nuclear Research of the Academy of Sciences of
Russia,\\ 117312, Moscow, Russia\\
\end{center}
\begin{center}
{\bf ABSTRACT}
\end{center}
\noindent
Considering   the results of recent distinguished analytical
calculations of the 5-loop single-fermion loop 
corrections  to the QED $\beta$-function we emphasize that to our 
point of view  it is 
important to perform their independent   cross-checks.  
We propose one of the ways of these  cross-check.
It  is based on the application of the original Crewther relation. 
We derive the new analytical 
expressions for the  $C_F^4\alpha_s^4$-contributions
to  the Bjorken polarized sum rule.
If results of  possible direct calculations will 
agree with the presented expression, then the appearance of 
$\zeta_3$-term in the 5-loop correction to the QED 
$\beta$-function and in  the $C_F^4\alpha_s^4$ contribution 
into  the $e^+e^-$ annihilation Adler function will get 
independent support and may be 
analysed  within the framework of the recently introduced 
concept of ``maximal 
transcendentality''.

 \vspace*{0.1cm}
\noindent
\\[3mm]
\vfill\eject

\setcounter{page}{1}
\pagestyle{plain}

\section{Introduction}

Quite recently the complicated  analytical expression  
for  the  non-singlet order $\alpha_s^4$  contribution  to the 
$e^+e^-$ annihilation Adler function    
\begin{equation}
\label{adlerNS}
D^{NS}(Q^2)=Q^2\int_0^{\infty}
\frac{R(s)}{(s+Q^2)^2}ds
=3\sum_F Q_F^2C_D^{NS}(a_s(Q^2))
=3\sum_F Q_F^2
\bigg[1+\sum_{n=1}^{n=4}
d_n^{NS}a_s^{n}\bigg]
\end{equation}
appeared in the literature    \cite{Baikov:2008jh}. 
Here $R(s)$   is the well-known $e^+e^-$ ratio, $Q_F$ are the quarks
charges,  $a_s=\alpha_s(Q^2)/\pi$ and  $\alpha_s(Q^2)$ is the 
$\rm{\overline{MS}}$-scheme QCD coupling constant, which obeys 
the property of asymptotic freedom  at large $Q^2$.  
The evaluation   of  $d_4^{NS}$  \cite{Baikov:2008jh} is    the third 
 step after   
analytical calculations of the $\alpha_s^2$ \cite{Chetyrkin:1979bj}
and $\alpha_s^3$ corrections \cite{Gorishnii:1990vf}
\cite{Surguladze:1990tg} 
to the Adler function of vector currents. The expression  for 
the $\alpha_s^3$-term   was confirmed 
later on by really independent calculation of  Ref. \cite{Chetyrkin:1996ez}.
However, the first theoretical argument in favour 
of the validity 
of the result of Ref. \cite{Gorishnii:1990vf} came from the 
foundation of Ref.\cite{Broadhurst:1993ru}, where  it was shown 
that the product of 
the order $\alpha_s^3$-expression for the $D^{NS}$-function 
and of the similar approximation  for the Bjorken polarized sum rule 
\cite{Larin:tj}  is leading to the   one-scale generalization 
of  the quark-parton Crewther relation 
\cite{Crewther:1972kn}. This generalized expression receives 
 extra term, proportional 
to the two-loop QCD $\beta$-function  \cite{Broadhurst:1993ru}.
The guess that 
this foundation will be correct in all orders of perturbation theory 
was made with caution in Ref. \cite{Broadhurst:1993ru}   
and at  more confidence level  in Ref. \cite{Gabadadze:1995ei}.
Moreover, extra arguments in favour of relating  this property  
to the effect of violation  of the conformal symmetry of 
massless theory of strong interactions by the  
terms, proportional to the factor $\beta(a_s)/a_s$,  were 
given in Ref.\cite{Gabadadze:1995ei} in momentum space. 
Later on this property  got more solid support  
after its   detailed proof, performed   in coordinate space 
\cite{Crewther:1997ux}.
In this letter  I will show, how  the application of the analog  
of the original Crewther relation \cite{Crewther:1972kn} 
may help to get deeper  understanding of the status of the 
5-loop QCD result of Ref. \cite{Baikov:2008jh} and of the part of its  
QED limit \cite{Baikov:2007}. Note, that both these 
analytical expressions  are giving rise to definite personal 
worries, which will be specified below. In view of this  it seems urgent  
to propose concrete ways of  their independent cross-check.

\section{Formulation of the problems}   

The result of Ref.  \cite{Baikov:2008jh},  namely  
Eq. (\ref{adlerNS}),  was presented  in the case of $SU(3)$ 
group only,  without singling out the corresponding Casimir operators
$C_F$ and $C_A$. This does not allow one  to study 
special theoretical  features of    $\alpha_s^4$-coefficients to both  
 $D^{NS}(Q^2)$ and to the photon vacuum polarization  constant $Z_{ph}$ 
in particular, which are manifesting themselves 
at the $\alpha_s^3$-level  in the case of $SU(N)$ group.    
Indeed, in Ref. \cite{Gorishnii:1990vf} it was observed, 
that  at the $\alpha_s^3$-level   $\zeta_3$-term, which appears 
in $Z_{ph}$ in QCD, is cancelling out 
in the case of $SU(N)$ gauge group with $C_A=C_F=Tf/2=N$, i.e. in the case 
of the concrete renormalization group constant of   
$SU(4)$ supersymmetric Yang-Mills theory,  studied in  detail 
at the  three-loop level 
in Ref. \cite{Avdeev:1980bh}. This observation 
gave the authors of Ref.  \cite{Gorishnii:1990kd} some  
additional theoretical arguments  in favour of the validity  of the part 
of the obtained in this work 4-loop  results. 
It will be highly 
desirable to get similar gentle support of the validity of 
5-loop QCD expression  of Ref.~\cite{Baikov:2008jh}.  
 
However, at present at this level 
there are extra unexplained theoretical questions.
Indeed, let us have a look to  the structure of   
interesting part of analytical result of
Ref.\cite{Baikov:2008jh}, namely to the perturbative 
expression for  the  single-fermion    
contribution to the QED $\beta$-function (which is proportional 
to the single-fermion QCD contribution to Eq.(\ref{adlerNS}).
Its 5-loop  expression  was presented in Ref.~\cite{Baikov:2007}   
and has the following form:
\begin{eqnarray}       
\label{beta{QED}}
\beta_{QED}^{[1]}&=&\frac{4}{3}{\rm{A}}+4{\rm A}^2-2{\rm A^3}-46{\rm{A^4}}
+\bigg(\frac{4157}{6}+128\zeta_3\bigg){\rm A}^5 \\ \nonumber
&=&\frac{4}{3}{\rm{A}}\times C_D^{NS}(\rm{A})
\end{eqnarray}
where ${\rm A}=\alpha/(4\pi)$ and $\alpha$ is the QED coupling constant.

It can be shown that the coefficients of Eq.(\ref{beta{QED}}) 
are scheme-independent (see e.g. \cite{Broadhurst:1992za}), 
at least in the schemes, not related 
to the  lattice regularization. This property is related 
to the conformal symmetry of the subsets of graphs, contributing 
to Eq.(2). In this limit the expansion parameter
$\rm{A}$ is not running and is simply the constant (it does not 
depend from  any scale). 
 The analytical structure of the 5-loop   result of Ref. 
\cite{Baikov:2007} differs from the previously known terms:
it contains  $\zeta_3$-term in the 5-loop coefficient.

Note, that  at the intermediate stages of calculations of the 3-loop 
correction  to Eq.(\ref{beta{QED}}) \cite{Rosner}, \cite{Bender:1976pw}
$\zeta_3$-terms  were  appearing,  but they  cancelled out  in the ultimate 
result.
Moreover, in Ref. \cite{Bender:1976pw} this feature was  
related to the  property of the conformal invariance of 
this part of QED $\beta$-function,     
though no proofs or references were given.

Next, in the process of evaluation  of the  4-loop term in 
Eq.(\ref{beta{QED}})  \cite{Gorishnii:1990kd}
the contributions with two  transcendentalities 
$\zeta_3$ and $\zeta_5$ appeared  at the intermediate stages 
of calculations, but these contributions cancelled   in the 
final result.

At the five-loop level one may expect, that 
$\zeta_3$, $\zeta_5$ and $\zeta_7$ should appear, 
but cancel down in the final result.
However, Eq.(2) demonstrate that for $\zeta_5$ and $\zeta_7$ 
this property is valid, while for $\zeta_3$ this is not the case!

Personally, I do not know any examples where the similar 
features, namely  the cancellations of higher transcendentalities,
but appearance of lower ones  in higher orders,  despite 
their cancellation at lower orders,  are manifesting themselves. 
I do not know
whether this observation  may be related to the un-proved property   
of ``maximal  transcendentality'', which at present is widely discussed 
while considering perturbative series for different quantities   
in the conformal invariant  $N=4$ 
SYM theory 
(see e.g. \cite{Kotikov:2006ts}, \cite{Drummond:2007aua}).
Thus we do not know whether the appearance of the transcendental 
term may be considered pro or contra the validity of the 
 results of Refs. \cite{Baikov:2007}, \cite{Baikov:2008jh}.

In any case, to clarify the status of this new feature 
of perturbative series in QED it is highly desirable  to get 
{\it independent calculational verification } of the results 
of Ref.  \cite{Baikov:2007},  \cite{Baikov:2008jh}. 

\newpage

\section{Proposed procedures of cross-checks}

The study  
of the prediction of the coefficient before $C_F^4\alpha_s^4$ contribution 
to the perturbative QCD term in the Bjorken sum rule of the polarized 
charged lepton- polarized nucleon deep-inelastic scattering
is one of the ways, which may allow  to understand  better 
the status of the results of Eq.(\ref{beta{QED}}).
 This 
sum rule can be defined as   
\begin{equation}
{\rm Bjp}(Q^2)=\int_0^1\bigg[g_1^{lp}(x,Q^2)-g_1^{ln}(x,Q^2)\bigg]dx=
\frac{1}{6}g_AC_{Bjp}(a_s)=\frac{1}{6}g_A\bigg[1+\sum_{n=1}^{n=4}
c_n a_s^n\bigg]
\end{equation}
Using the conformal-invariant (c-i)  limit of the generalized Crewther 
relation, discover in Ref. \cite{Broadhurst:1993ru}, it is possible 
to write-down the following relation 
\begin{equation}
C_{Bjp}(a_s(Q^2))C_D^{NS}(a_s(Q^2))|_{c-i}=1~~~,
\label{17}
\end{equation}
It follows from application of operator product expansion method for the  
three-point function of axial-vector-vector non-singlet quark currents 
in the momentum space  \cite{Gabadadze:1995ei} (for more details 
see \cite{Kataev:1996ce}) and is reproducing   
original Crewther relation, obtained from  
the coordinate space considerations of Ref. \cite{Crewther:1972kn}   and  
Ref. \cite{Adler:1973kz} as well.  
Note also that 
Eq.(\ref{17}) differs from the one, derived  in Ref. \cite{Brodsky:1995tb} 
(for the related analysis see  Ref.\cite{Rathsman:1996jk}).
Indeed,  in Eq.(\ref{17}) the coupling constant $a_s$ is  
{\bf scale independent} and is  defined in the Euclidean region.

Taking into account  the results of previous QCD calculations 
and generalizing  5-loop result of Ref.\cite{Baikov:2007} 
to the case of QCD in the conformal invariant 
limit, one has
\begin{equation}
\label{Baikov2007}
C_D^{NS}(a_s)=\bigg[1+\frac{3}{4}{\rm C_F}a_s-\frac{3}{32}
{\rm C_F^2}a_s^2-\frac{69}{128}{\rm C_F^3}a_s^3+\bigg(\frac{4157}{2048}
+\frac{3}{8}\zeta_3\bigg){\rm C_F^4}a_s^4 \bigg]~~~.
\end{equation}
where  
${\rm C_F}=(N^2-1)/(2N)$ in the case of  $SU(N)$ gauge group.
Using now Eq.(\ref{17})  we get   
scheme-independent  contributions to the Bjorken polarized sum rule,
which include  two new order   $\alpha_s^4$ terms  
 \footnote{It is possible to show that in the conformal invariant 
limit logarithmic QCD   contributions  to the 
 Gross-Llewellyn Smith sum rule coincide with the ones for the Bjorken 
polarized sum rule in all orders of perturbation 
theory, see e.g. \cite{Kataev:1996ce}.}:   
\begin{equation}
\label{results}
C_{Bjp}(a_s)=1-\frac{3}{4}{\rm C_F}a_s
+\frac{21}{32}{\rm C_F^2}a_s^2 -\frac{3}{128}{\rm C_F^3}a_s^3
-\bigg(\frac{4823}{2048}+\frac{3}{8}\zeta_3\bigg){\rm C_F^4}a_s^4
\end{equation}
The coefficients of  order $a_s$, $a_s^2$ and $a_s^3$-terms  
are in agreement with the result of explicit calculations, performed 
in Refs.\cite{Kodaira:1979ib},~\cite{Gorishnii:1985xm} and~ \cite{Larin:tj}
respectively. It should be also mentioned that the similar consideration 
was performed previously in Ref. \cite{Adler:1973kz} at the level of 
$a_s$ corrections, but the $a_s^2$-term was not predicted there.

The direct   evaluation   of the {\bf predicted} $a_s^4$
coefficient  may be rather useful for the  independent 
cross-check   of the  QED results of  Ref.\cite{Baikov:2007} and thus of  
the related part of the QCD expression from 
Ref.  \cite{Baikov:2008jh}. This evaluation 
should clarify whether $\zeta_3$ term is appearing in the $a_s^4$ 
correction to $C_{Bjp}(a_s)$ or not. This will give the most decisive 
argument pro or contra the validity  of the $\alpha_s^4$  results of 
Eq.~(\ref{Baikov2007}), which are following from the ones of 
Eq.(\ref{beta{QED}}), presented in \cite{Baikov:2007}.  

Note, that 
there are also at least two other possibilities 
for the cross-check of the result of Eq.(\ref{beta{QED}}).
The first one is related to the extension to 5-loops  of
Dyson-Shwinger-Johnson motivated analysis, performed by Broadhurst 
\cite{Broadhurst:1999zi}
at the 4-loop level. 
The 5-loop extension of the work of Ref.\cite{Broadhurst:1999zi},
based on   the calculations of definite  
5-loop anomalous dimensions  in QED from the    4-loop finite  
scheme-independent integrals,  
should demonstrate the cancellation of  $\zeta_5$ and $\zeta_7$ terms 
and clarify whether  $\zeta_3$- contribution is appearing or not.

Another way  for checking the result 
 of  Eq.(\ref{beta{QED}}) may be  based on the generalization  of 
the Background Field Method to the case of   5-loop QED calculations.
Note, however, that  
up to now this method was directly used at the 3-loop level only  \cite{Bornsen:2002hh}.

\section{Conclusion}

In this letter we address the question on the available at present 
possibilities of independent cross-checks of the part 
of the result of Ref. \cite{Baikov:2008jh}. To our point 
of view the most decisive and urgent  test may come 
from evaluation of the coefficient of   $C_F^4\alpha_s^4$ 
contribution to the Bjorken sum rule, which may present additional 
arguments pro or contra the appearance  of $\zeta_3$-term 
in the 5-loop perturbative correction 
of  one-fermion loop contribution 
into the QED $\beta$-function.

This work is based on the talk at the 15th  International 
Seminar ``Quarks-2008'', May 23-29, 2008, Sergiev Posad,  Russia
(for its preliminary write-up see \cite{Kataev:2008nc}).
I am grateful to its participants for rather useful discussions.  
The work is supported by  RFBR Grants 
N 08-01-00686-a and 06-02-16659-a.

\end{document}